\documentclass{aa}  

\usepackage{graphicx}
\usepackage{txfonts}
\usepackage[colorlinks=true,linkcolor=blue,citecolor=blue, urlcolor=blue]{hyperref}
\usepackage{units}
\usepackage{xcolor}
\usepackage{amsmath}
\usepackage{lscape}
\usepackage{nicefrac}
\usepackage[utf8]{inputenc}
\usepackage{caption}
\usepackage{subcaption}

\begin{document}

   \title{Gaia DR3 Variable White Dwarfs vetted by ZTF}


   \author{Timour Jestin\inst{1}
   \and 
   \foreignlanguage{vietnamese}{Thịnh Hữu Nguyễn}\inst{2,3}
   \and
   Laurent Eyer\inst{1}
   \and
   Lorenzo Rimoldini\inst{4}
   \and
   Ashish Mahabal\inst{5}
   \and
   Marc Audard\inst{1}
   \and
   Pedro García-Lario\inst{6}
   \and
   Panagiotis Gavras\inst{7}
   \and
   Krzysztof Nienartowicz\inst{4,9}
   }

   \institute{Department of Astronomy, University of Geneva, Chemin Pegasi 51, 1290 Versoix, Switzerland
   \and 
   Department of Astronomy, University of Illinois Urbana-Champaign, 1002 W. Green Street, Urbana, IL, 61801, USA
   \and 
   Department of Astrophysics and Planetary Sciences, Villanova University, 800 E. Lancaster Avenue, Villanova, PA, 19085, USA
   \and
   Department of Astronomy, University of Geneva, Chemin d'Ecogia 16, 1290 Versoix, Switzerland
   \and 
   Division of Physics, Mathematics, and Astronomy, California Institute of Technology, Pasadena, CA 91125, USA
   \and
   European Space Astronomy Centre (ESA/ESAC), Villanueva de la Canada, 28692 Madrid, Spain
   \and 
   Starion for European Space Agency (ESA), Camino bajo del Castillo, s/n, Urbanizacion Villafranca del Castillo\\ Villanueva de la 21 Cañada, 28692 Madrid, Spain
   \and
   Sednai Sarl, Rue des Marbriers 4, 1204 Geneva, Switzerland
   }
   \date{\today}

 
  \abstract
   {
    The publications of \textit{Gaia} DR2 and DR3 have brought major improvements in stellar astrometry and photometry, particularly regarding the description of the white dwarf sequence. In particular, \textit{Gaia} DR2 enabled the detection of variability in white dwarfs based solely on averaged astrometric and photometric quantities, i.e. the astrometric 5 parameters (positions, proper motion, and parallax) and general photometric properties in the G, BP and RP bands (mean, standard deviation and number of measurements).
    }
    {
    We identify and classify variable white dwarfs using \textit{Gaia} DR3 data and Zwicky Transient Facility (ZTF) DR23 observations. The objective is to construct a catalogue of pulsating white dwarf candidates with robust selection criteria.
    }
    {
    We define a new sample of candidate variable white dwarfs using \textit{Gaia} DR3 astrometric and photometric data. We cross-match this sample with the ZTF DR23 catalogue and apply a multiband Lomb-Scargle periodogram analysis to detect periodic variability. We then use the \texttt{OPTICS} unsupervised clustering algorithm to group and classify the confirmed periodic stars.
    }
    {
    We identify 1423 variable white dwarf candidates from \textit{Gaia} DR3, with 864 having ZTF time series. Among these, 141~present significant periodicity. Using unsupervised clustering techniques, we classify these objects into known categories, including ZZ\,Ceti stars, GW\,Vir stars, V777 Her stars, and white dwarf–main sequence (WD–MS) binaries. Our analysis reveals several confirmed periodic stars, including three ZZ Ceti stars, 15 GW Vir stars, one V777 Her, and 24 WD–MS binaries. 
    Furthermore, our work reveals a significant population of potentially variable stars, even though without detected periodicity.
    Finally, several variables show unusual properties and are highlighted for follow-up studies.
    }
    {
    Our catalogue of candidate variable white dwarfs includes variability status, periodicity, and classification information for the 864 sources with ZTF time-series photometry, 519 of them newly identified, including 67 new periodic stars. It is publicly available through the VizieR database to support future investigations of white dwarf variability.
    }


   \maketitle

\section{Introduction} \label{sec:intro}
We are in an epoch of enormous data growth in optical astronomy. The challenge lies not only in the increase in data volume, but also in the variety of survey properties and data types. Among all astronomical surveys, the \textit{Gaia} mission \citep{PrustiEtal2016}, a cornerstone of the ESA science program launched in late 2013, is unique. From a single space-based platform, it delivered repeated astrometric, photometric, spectrophotometric, and spectroscopic measurements over the entire celestial sphere for more than 10 years.

The \textit{Gaia} DPAC consortium \citep{MignardEtal2008} has taken care of the data reduction, calibration and analysis of the \textit{Gaia} data and has also prepared the data products to be published by ESA in a stepwise manner. 
This approach aims to improve the data reduction, lower random and systematic errors in all data products, and gradually increase the number of published sources and data products. The same iterative approach was followed for publishing the \textit{Gaia} photometric time series. 
For the first \textit{Gaia} Data Release \citep[September 2016;][]{2016A&A...595A...2G, EyerEtal2017}, a selection of 3,194 variable stars near the South Ecliptic pole that partially covers the Large Magellanic Cloud was published with their G-band time series. 
With \textit{Gaia} DR2 \citep[April 2018;][]{BrownEtal2018}, more than half a million variable objects were published with their \textit{G}, integrated \textit{BP}, and \textit{RP} time series. 
\textit{Gaia} DR3 \citep{GaiaDR3_Summary, GaiaCollaboration2020, EyerEtal2023, Riello2020} then released the \textit{G}, \textit{BP} and \textit{RP} epoch photometry of some 11.7 million sources.
This represents one of the largest published catalogues for many variability types - such as RR Lyrae stars, large-amplitude long-period variable stars, and $\delta$\,Scuti stars.
Furthermore, a performance verification paper \citep[][published as companion to \textit{Gaia} DR2]{GaiaEyer2019} describes the variability across the observational Hertzsprung-Russell (HR) diagram, notably including distinctive variability features in the white dwarf sequence.

Within \textit{Gaia} data, the published averaged statistics can help infer the variability of stars whose time series have not been released \citep[for example,][]{BelokurovEtal2017,MowlaviEtal20202, GuidryEtal2021, MaizApellaniz2023}. 
Using the relationship between the uncertainty of the mean and its associated standard deviation, corrected for an estimation of the instrumental and photon noises \citep{Eyer1998}, it is possible to study the variability of certain HR regions with a comprehensive approach. 
Notably, \cite{EyerEtAl2020} explored the variability of the white dwarfs identified by \cite{GentileFusilloEtal2019}. 

The aim of this article is to study variable white dwarfs, which are selected from the \cite{GentileFusilloEtAl2021} white dwarf catalogue, using \textit{Gaia} DR3 and the Zwicky Transient Facility (ZTF) DR23 data \citep{BellmEtal2019, masci_zwicky_2018} as well as to compare the candidates with the literature. 
We aim to present a list of pulsating white dwarf candidates with adjusted criteria employing \textit{Gaia} DR3 measurements while also considering other variable types.
A systematic search for variability in white dwarfs is particularly valuable, as it can reveal insights into their physical properties, especially for those located in specific regions in the HR diagram, such as the crystallisation sequence.

The article is structured as follows. First, we present some general properties of the \textit{Gaia} and ZTF surveys in Section~\ref{Sect:Data}. In Section~\ref{Sect:Gaia_EDR3_candidates}, we show the criteria used to select and refine eligible variable white dwarf candidates from \textit{Gaia} DR3. After obtaining the list of candidates, we describe our procedure to cross-match with ZTF data and extract the frequency from the time series in Section~\ref{sec:period_search}. Then, Section~\ref{Sect:pulsating_white_dwarf_candidates} presents our selection of pulsating white dwarf candidates. 
We also classify these candidates using unsupervised clustering, and cross-match them with known stars from the literature. 



\section{Data}
\label{Sect:Data}

\begin{figure*}[h]
    \centering
    \includegraphics[width=0.7\textwidth]{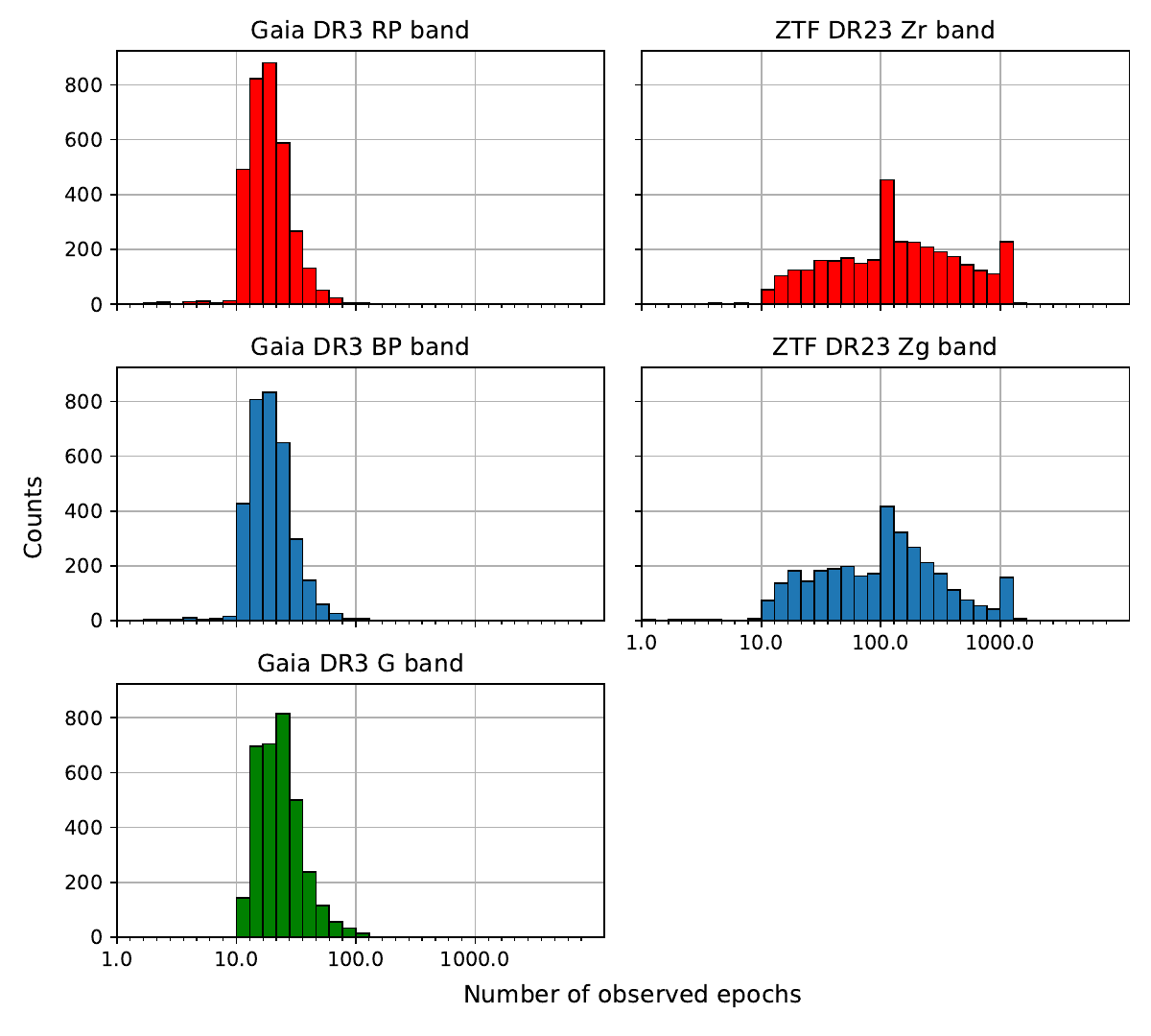}
    \caption{The histograms display the number of observed epochs in each band for each survey. We randomly choose 3000 white dwarf candidates from the \cite{GentileFusilloEtAl2021} catalogue and extract their number of measurements in the two surveys. On average, ZTF~DR23 has a greater number of observations per star than the number of FoV transits per star from \textit{Gaia} DR3.}
    \label{fig:numobshistogram}
\end{figure*}

In this study, we use \textit{Gaia} DR3 and ZTF DR23 data obtained from their respective public archives\footnote{\url{https://gea.esac.esa.int/archive/}}$^,$\footnote{\url{https://irsa.ipac.caltech.edu/Missions/ztf.html}}.
We use the \textit{Gaia} photometry \citep{EvansEtal2018,Riello2020} and astrometry \citep{LindegrenEtal2018,Lindegren2020} measurements to determine our targets' locations in the Hertzsprung-Russell (HR) diagram. The ZTF time series in the \textit{g} and \textit{r} bands (referred to as Zg and Zr, respectively) are then used to determine potential periodic stars among our candidates.

We study the general properties of the \textit{Gaia} and ZTF observations using a random sample of about 3000 white dwarf candidates from \cite{GentileFusilloEtAl2021}'s \textit{Gaia} DR3 catalogue that have ZTF time series.
Figure~\ref{fig:numobshistogram} displays the number of epochs observed in each band of \textit{Gaia} DR3 and ZTF DR23. These are defined as the number of Field-of-View (FoV) transits for \textit{Gaia}, and the number of observations for ZTF. According to the figure, the sources have a lower number of observed epochs in \textit{Gaia} DR3 compared to ZTF DR23, with a median number of observed epochs of 41, 33, and 34, in \textit{Gaia}'s G, RP, and BP bands respectively, while these reach 129 in the Zg band and 190 in the Zr band.
While the ZTF data yields higher noise per measurement than \textit{Gaia}, the higher number of measurements should allow for comparable error levels on the mean magnitudes.

\begin{figure}
    \centering
    \includegraphics[width=0.95\columnwidth]{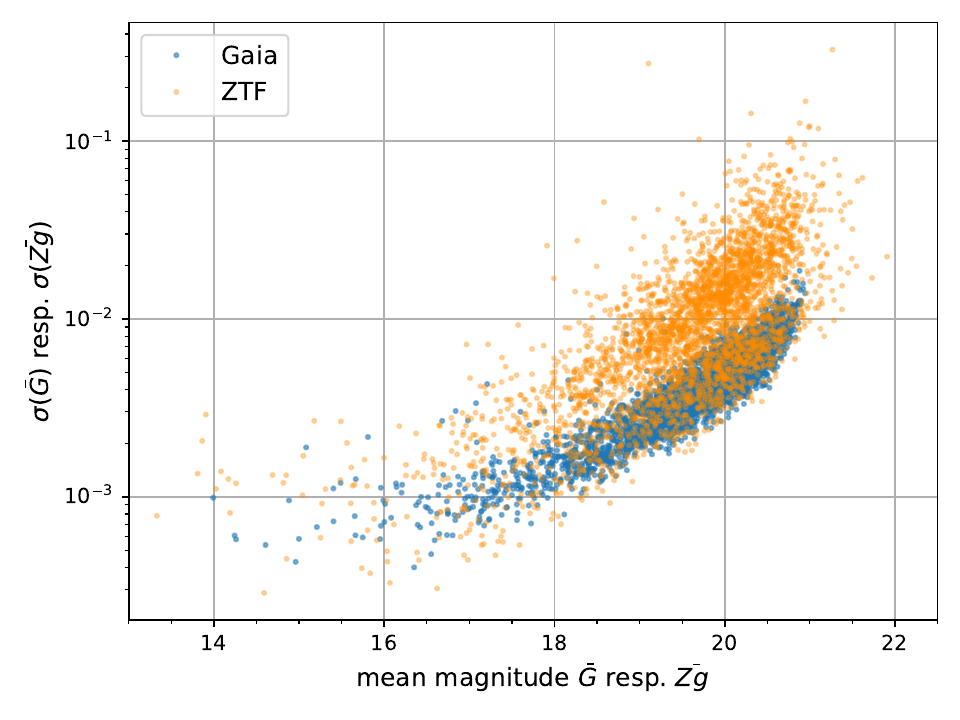}
    \caption{Comparison of the error of the mean magnitude $\sigma$ as a function of the mean magnitude between \textit{Gaia} DR3 (G band) and ZTF~DR23 time series (Zg band) for individual sources. Around 3000 white dwarf candidates are chosen randomly from the \cite{GentileFusilloEtAl2021} catalogue and the standard deviations of their mean magnitude are computed. As \textit{Gaia}~DR3 yields very high precision for individual measurements and ZTF~DR23 has many measurements, both surveys present low noise for their mean magnitudes.}
    \label{fig:comparison_precision}
\end{figure}

We then compare the standard deviation of the time series $\sigma(G)$ and the error of the mean magnitude $\sigma(\bar{G})$ between the two datasets. For \textit{Gaia} data, these are estimated using the number of observations per CCD (\texttt{phot\_g\_n\_obs}) and the photometric mean flux over its error (\texttt{phot\_g\_mean\_flux\_over\_error}):

\begin{equation}
    \sigma(G) = \frac{2.5}{\ln10}\cdot\frac{\sqrt{\texttt{phot\_g\_n\_obs}}}{\texttt{phot\_g\_mean\_flux\_over\_error}},
    \label{eq:GaiaStandardDeviationCalculation}
\end{equation}

\begin{equation}
    \centering
    \sigma(\bar{G}) = \frac{2.5}{\ln10}\cdot\frac{1}{\texttt{phot\_g\_mean\_flux\_over\_error}}.
    \label{eq:GaiaStandardDeviationofMeanCalculation}
\end{equation}
For the ZTF data, $\sigma(\bar{Zg})$ is calculated by dividing the time series standard deviation by the square root of the number of measurements in the Zg band.
Figure~\ref{fig:comparison_precision} displays the error of the mean magnitude as a function of the mean magnitude in the G resp. Zg bands, computed on our previous star sample.
It demonstrates that both \textit{Gaia} and ZTF have comparably low noise (below 10 millimag, for $G<18$) and confirms that ZTF data can be used to detect periodic signals from \textit{Gaia}'s variable white dwarf candidates even when photometric time series are unavailable in \textit{Gaia}. 
Moreover, ground-based surveys like ZTF typically have a spectral window with peaks at multiples of one cycle per day. In the case of \textit{Gaia}, the period of rotation is six hours, inducing peaks at multiples of four cycles per day. This difference in survey cadence can lift aliasing ambiguities when combining the two surveys.

\section{Selecting \textit{Gaia}~DR3 variable white dwarf candidates}
\label{Sect:Gaia_EDR3_candidates}

\subsection{Candidacy criteria}\label{sub:candidates}

We obtained our variable white dwarf candidates from \cite{GentileFusilloEtAl2021}'s \textit{Gaia} DR3 white dwarf catalogue. Even though these stars have already been selected using strict quality criteria on both photometric and astrometric flags, we decided to take stricter and more conservative restrictions to avoid contamination for our subsequent selection. 
In particular, dimmer sources (as shown in Figure~\ref{fig:comparison_precision}) and sources with a high colour index likely exhibit large photometric variability due to noise rather than intrinsic variability. 
Thus, we impose our own restrictions on these parameters:
\begin{equation}
\begin{aligned}
    & P_{WD} \geq 0.75, \\
    & \tt{phot\_g\_mean\_mag} < 18, \\
    & \tt{phot\_bp\_mean\_mag} - \tt{phot\_rp\_mean\_mag} \leq 2, \\
    & \tt{astrometric\_excess\_noise\_sig} \leq 15, \\
\end{aligned}
\end{equation}
where $P_{WD}$ is the probability of being a white dwarf calculated by \cite{GentileFusilloEtAl2021} and the rest are the \textit{Gaia} parameters. 
We also note that these restrictions are empirical and are selected based on the cleanliness of the resulting over-density regions later on. 
These limits may exclude true variable sources but are essential to ensure the purity of the selection. 
This filtering reduces the selection from $\sim$1.3 million stars in the initial catalogue to 22264 sources.

\begin{figure}
    \centering
    \includegraphics[width=0.95\columnwidth]{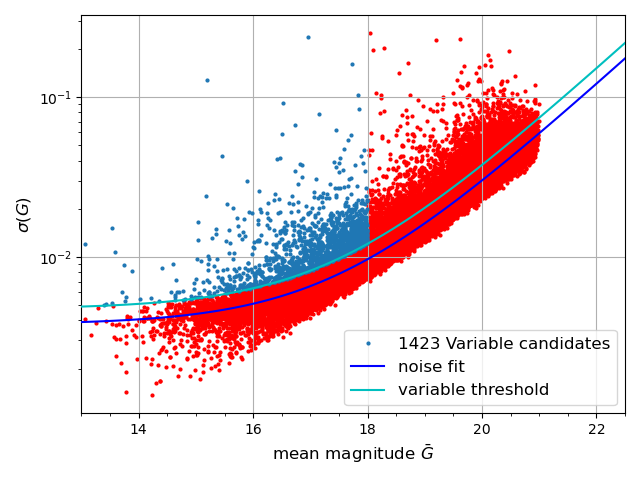}
    \caption{The estimated standard deviation of the \textit{G}-band per CCD time series, Equation~(\ref{eq:GaiaStandardDeviationCalculation}), plotted as a function of the mean magnitude $\bar{G}$. To select the variable white dwarf candidates, we use a threshold that is 25\% higher than the computed noise function. Using this selection method, we obtain 1423 variable dwarf candidates from \textit{Gaia} DR3 data.}
    \label{fig:variable_selection}
\end{figure}

Next, we assess the variability of our remaining sources.
Figure~\ref{fig:variable_selection} displays the \textit{G}-band estimated standard deviation of the time series per CCD (from Equation~(\ref{eq:GaiaStandardDeviationCalculation})) as a function of the mean \textit{Gaia} G-band magnitude, $\bar{G}$.
The relationship between noise and mean magnitude is fitted as an exponential function $f(\bar{G})=\exp(a\cdot\bar{G}+b)+c$ using SciPy's \texttt{curve\_fit} function \citep{virtanen_scipy_2020}. This yields:
\begin{equation}
\begin{aligned}
    \sigma(G; \bar{G}) = & \exp\left((7.48\pm0.01)\cdot10^{-1}\cdot\bar{G} - 18.597\pm0.025\right) \\ 
    & + (3.8\pm0.1)\cdot10^{-3} \\
\end{aligned}
\label{eq:noise_threshold}
\end{equation}
With the notation $\sigma(G; \bar{G})\equiv\sigma(G)$ used for clarity.
We then define a threshold to define our candidate variable stars as: 
\begin{equation}
    \mbox{variability threshold}=1.25\cdot\sigma(G;\bar{G})
\end{equation}
i.e. a 25\% increase of the required $\sigma(G)$ compared to its average value, reducing our selection to 1423 variable white dwarf candidates. 

This method, based on \cite{EyerEtAl2020}, is similar to \cite{GuidryEtal2021}'s "Gaia Variability Metric", where they select sources with the 1\% highest $\sigma(G)$ values after magnitude de-trending (done using an exponential fitting).
However, in our work, we take into account the effects of magnitude on the $\sigma(G)$ metric using the "variability threshold".
Furthermore, the value of the "variability threshold" selects the top 6\% most variable sources, rather than the top 1\%.
This extension of the number of possible targets is a balancing act to include a higher proportion of potential variable sources 
without significantly increasing the computational costs.
Furthermore, unlike the work of \cite{GuidryEtal2021}, we rely exclusively on this \textit{Gaia}-based variability metric and use ZTF data as follow-up, restricting our final study to sources within the ZTF catalogue.

\begin{figure}
    \centering
    \includegraphics[width=0.95\columnwidth]{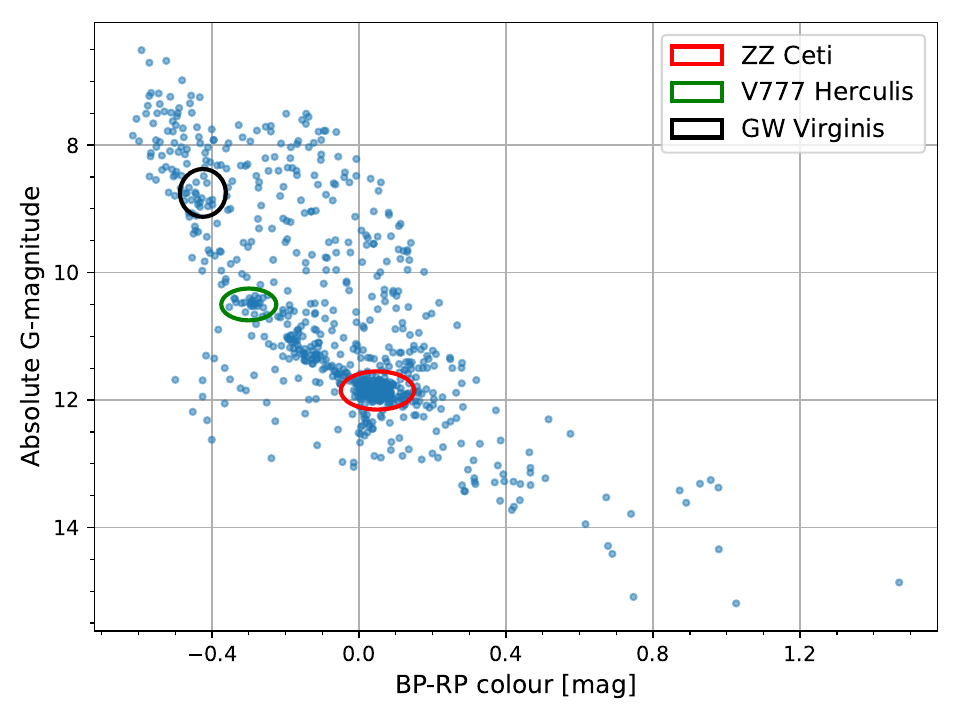}
    \caption{The HR diagram of our selection of variable white dwarf candidates from Figure~\ref{fig:variable_selection}. 
    The potential ZZ Ceti, V777 Herculis and GW Virginis clumps are outlined by ellipses.
    We use the mean magnitude in the \textit{Gaia} \textit{G}, \textit{BP}, and \textit{RP} bands.}
    \label{fig:variable_candidates}
\end{figure}

We present an HR diagram of our variable candidates in Figure~\ref{fig:variable_candidates}.
Along the white dwarf sequence, there are noticeable clump, notably the ZZ~Ceti clump ($\mbox{BP-RP}\in[0,0.1]$, $\mbox{absolute G}\in[11.6,12.1]$), a potential V777~Her clump ($\mbox{BP-RP}\in[-0.35,-0.25]$, $\mbox{absolute G}\in[10,11]$), and potential GW~Vir stars ($\mbox{BP-RP}\in[-0.45,-0.4]$, $\mbox{absolute G}\in[8,9]$).

\subsection{Robustness of the selection and false positives}\label{subsec:robust}

\begin{figure}
    \begin{subfigure}{0.95\columnwidth}
    \centering
    \includegraphics[width=\linewidth]{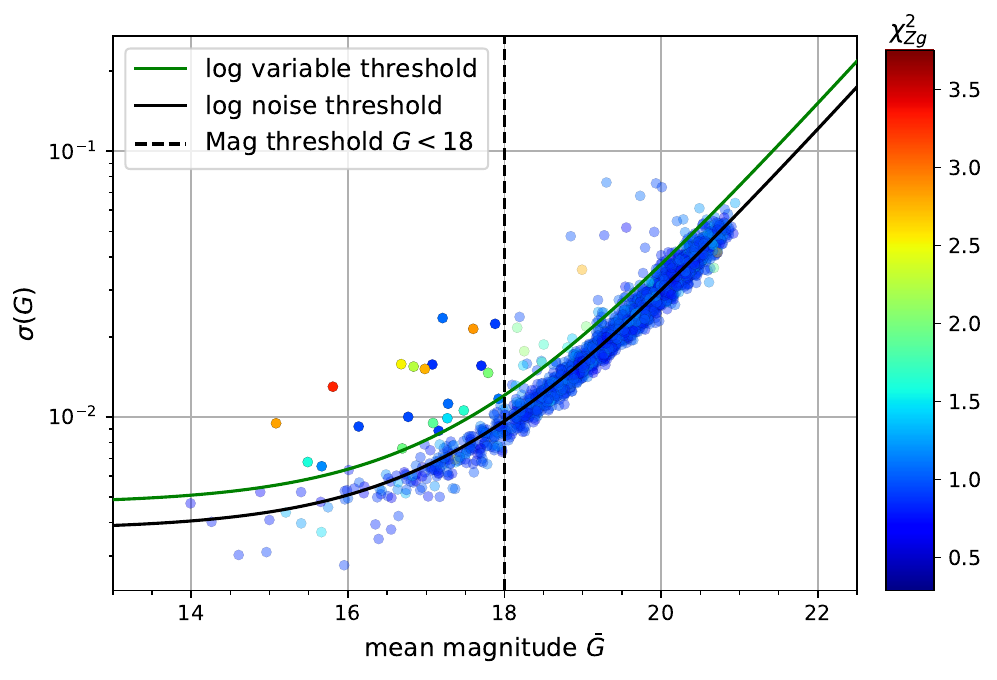}
    \caption{}
    \end{subfigure}
    \begin{subfigure}{0.95\columnwidth}
    \centering
    \includegraphics[width=\linewidth]{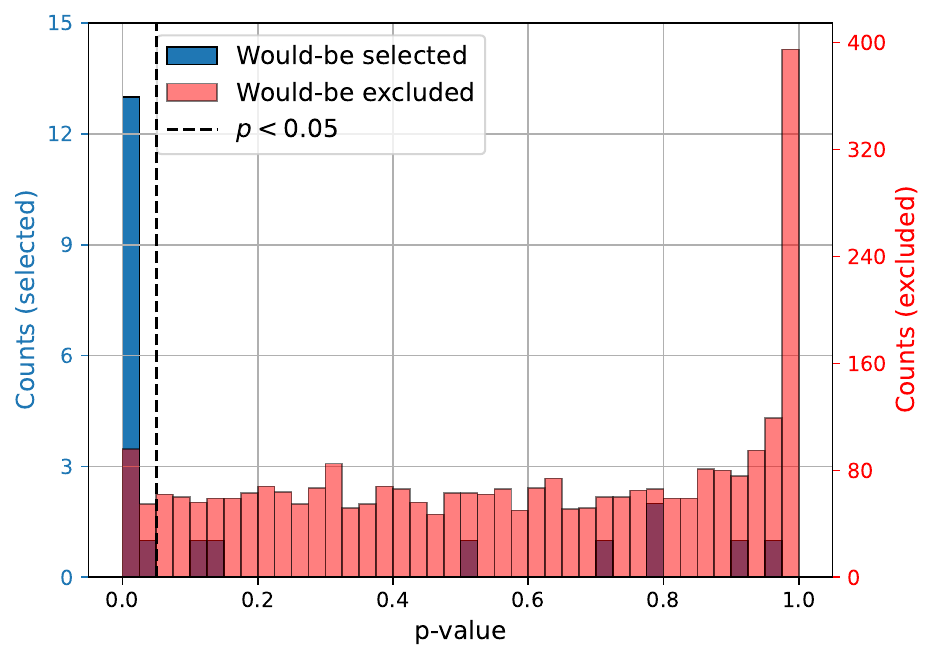}
    \caption{}
    \end{subfigure}
    \caption{(a) Reduced Zg-band $\chi^2$ value for the 3000 sample sources as function of their mean \textit{Gaia} $G$-band magnitude and $G$-band standard deviation. The would-be excluded sources are displayed with a high transparency.
    (b) Histogram of the corresponding Zg-band p-values for both the would-be selected and would-be excluded sources. We note increase in the p-value approaching the last bin, suggesting that uncertainties in the $\chi^2$ calculation are overestimated relative to their true values. }
    \label{fig:chi2_test}
\end{figure}

We stress again that our selection criteria are defined empirically based on the cleanliness of the resulting over-density regions visible in Figure~\ref{fig:variable_candidates}.
To verify the robustness of this selection, we perform Pearson's $\chi^2$ test on the Zg magnitudes of the 3000 sample sources presented in Section~\ref{Sect:Data}. 
We divide our sample in two categories, the "would-be selected" and the "would-be excluded", by applying our selection criteria (including the variability threshold). 
The reduced $\chi^2$ and respective p-values of each source the sample are presented in Figure~\ref{fig:chi2_test}.
By using the usual confidence level $p<0.05$, 64\% of the sources in the "would-be selected" category are variable, while the overall proportion of variable sources with magnitude $G<18$ identified using Pearson's $\chi^2$ test drops to 14\%.
Additionally, this allows us to compute an expected false positive rate of our selection method to $FPR=36\%$.
Finally, our method selects 45\% of the sources at $G<18$ identified as variable by the Pearson $\chi^2$ test.

\subsection{Comparison with \textit{Gaia} DR3 variability classifier}

We compare our sample of candidate variable white dwarfs with the \textit{Gaia} DR3 Machine Learning variability classification \citep{EyerEtal2023, RimoldiniEtal2023}.
This comparison then allows us to validate our candidacy criteria and can reveal possible missing sources in the \textit{Gaia} classification algorithm.
This yields: 994 (69.9\%) unclassified candidates, 408 (28.7\%) classified as white dwarfs, 8 as cataclysmic variables, 6 as eclipsing binaries, 5 as short time-scale variable stars and, most surprisingly, two RR Lyrae stars (Gaia DR3 3345661467822106624 and Gaia DR3 6555925496084361344) These two stars are not only classified as RR Lyrae but are also confirmed in the RR Lyrae catalogue by \cite{ClementiniEtAl2023}.
These are misclassifications of the \textit{Gaia} machine-learning classification and an erroneous confirmation by \cite{ClementiniEtAl2023}, as they are not located in the horizontal branch, as would be expected for such stars.


\section{Period search with ZTF data}
\label{sec:period_search}

\subsection{Evaluating the ZTF cross-match query}
\begin{figure}[h]
    \centering
    \begin{subfigure}{0.9\columnwidth}
    \centering
    \includegraphics[width=\linewidth]{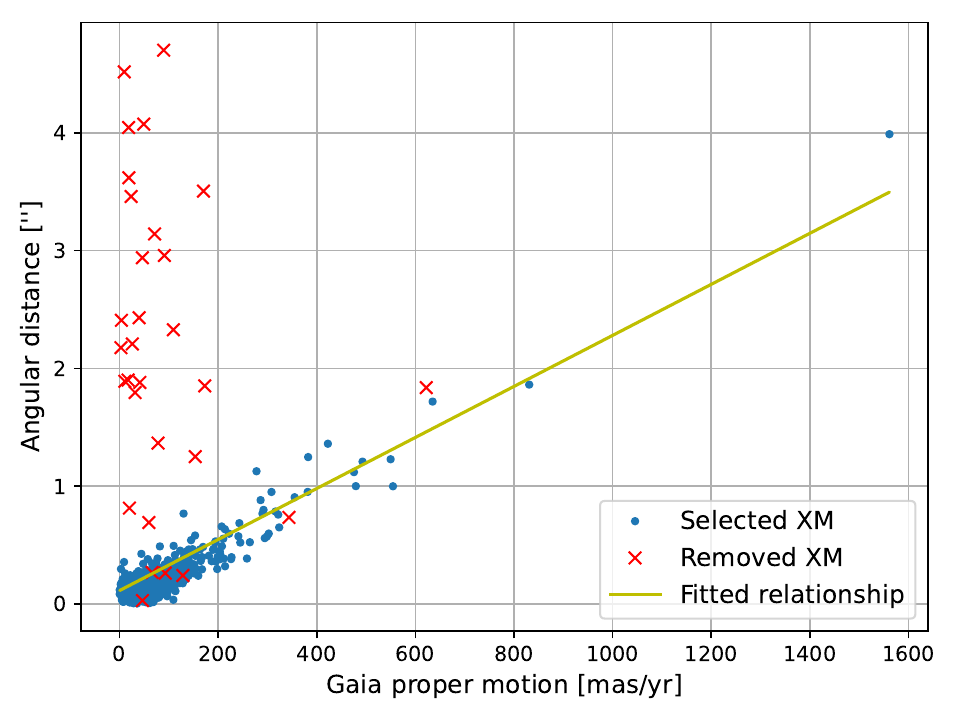}
    \caption{\textit{Gaia} proper motion vs angular distance.}
    \label{fig:outliers_angular_dist_pm}
    \end{subfigure}
    \hfill
    \begin{subfigure}{0.9\columnwidth}
    \centering
    \includegraphics[width=\linewidth]{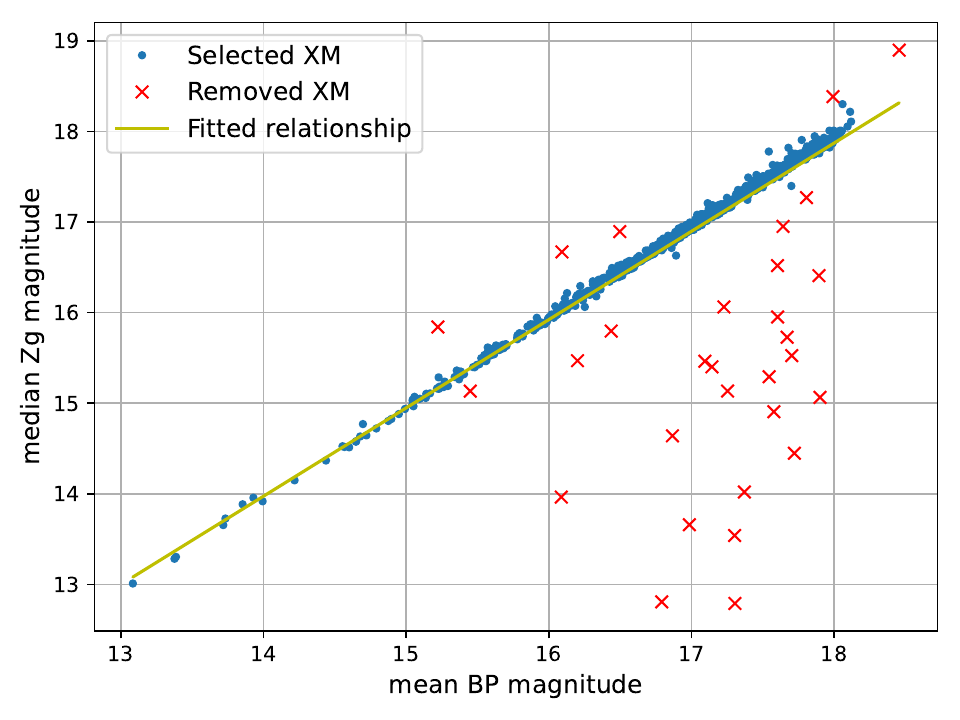}
    \caption{\textit{Gaia} BP mag vs Zg mag.}
    \label{fig:outliers_g_mag}
    \end{subfigure}
    \hfill
    \begin{subfigure}{0.9\columnwidth}
    \centering
    \includegraphics[width=\linewidth]{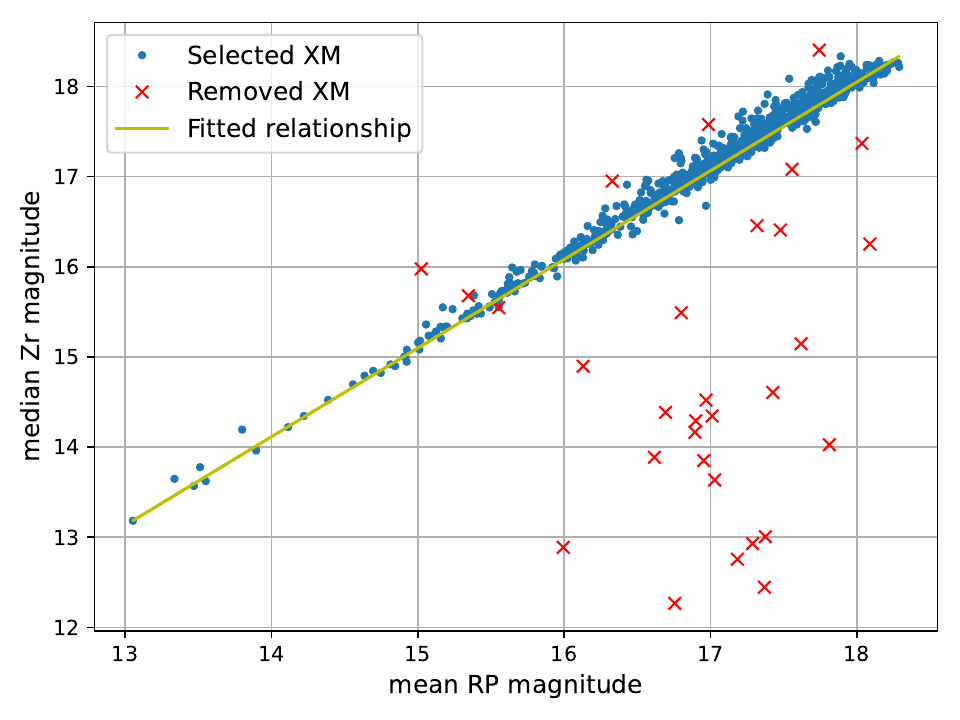}
    \caption{\textit{Gaia} RP mag vs Zr mag.}
    \label{fig:outliers_r_mag}
    \end{subfigure}
    \hfill
    \caption{Inconsistent cross-matches of the ZTF query are detected and removed by comparing the astrometric (subplot (a)) and photometric measurements (subplots (b) and (c)) between \textit{Gaia} and ZTF in the multi-dimensional space. This process removes 30 cross-matches out of the initial 894.}
    \label{fig:verifying_crossmatch}
\end{figure}

ZTF data are cross-matched with \textit{Gaia} by spatial constraint with a cone search of radius 10 arcsec. Each cross-match must have time series in both bands (Zg and Zr) and have at least 20 measurements per band. If multiple objects match the query constraint, the closest positional match will be selected. This selection is enabled automatically by the \textit{One to One Match} option in the ZTF archive. Among our 1423 variable white dwarf candidates, there are 894 stars with ZTF data available in both bands. However, wrong cross-matches can still happen, and additional steps to check the query are necessary. To detect potential erroneous cross-matches, we examine each data point's Euclidean distance between the normalized residuals of three linear regressions (done over the 894 stars with ZTF data available): the angular distance between \textit{Gaia} and ZTF coordinates versus \textit{Gaia} proper motion, the Zg mag versus \textit{Gaia} BP mag, and the Zr mag versus \textit{Gaia} RP mag. The fitted lines are fitted using SciPy's \texttt{curve\_fit}, and shown in Figure~\ref{fig:verifying_crossmatch}. The computed relationships are:
\begin{equation}
\begin{aligned}
    &\hat{\theta} = (2.168\pm0.160)\cdot10^{-3}\mu + (112.6\pm18.0)\cdot10^{-3}, \\
    &\hat{Zg} =  (0.974\pm0.016)\cdot BP+ (0.337\pm0.266), \\
    &\hat{Zr} =  (0.983\pm0.021)\cdot RP + (0.346\pm0.352)
\end{aligned}
\label{eq:least_square_lines}
\end{equation}
in which $\theta$ is the angular distance (in arcseconds) between \textit{Gaia} and ZTF coordinates, $\mu$ is the \textit{Gaia} DR3 proper motion (in milli-arcseconds per year), $Zg$ and $Zr$ are the ZTF magnitude measurements, and $BP$ and $RP$ are the \textit{Gaia} magnitude measurements. 

\begin{figure}
    \centering
    \includegraphics[width=0.95\columnwidth]{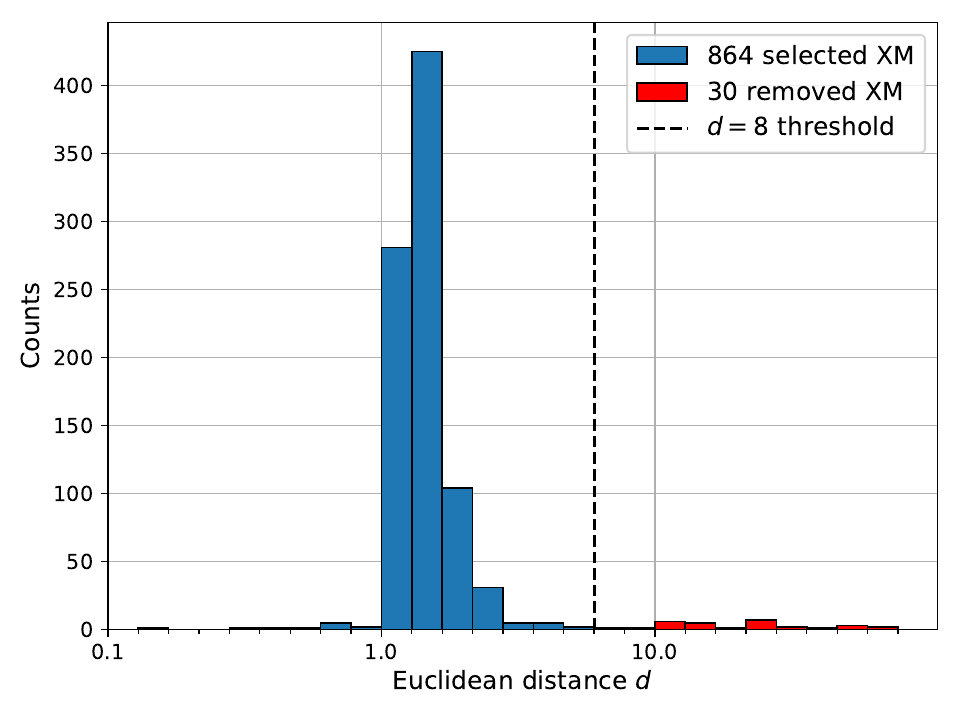}
    \caption{The distribution of the Euclidean distance - as calculated in Equation~(\ref{eq:Euclidean_distance}) - of all our cross-match query. Good cross-matches should have $d$ closer to 0. We remove those with $d$ larger than 8.}
    \label{fig:distribution_d}
\end{figure}

The residual of each variable with respect to its linear fit is normalized by the IQR of the residual's distribution. The IQR value is chosen over the standard deviation to avoid the influence of outliers, which are numerous in all three relationships (as shown in Figure~\ref{fig:verifying_crossmatch}). Including the normalization coefficients, the Euclidean distance of each data point is:
\begin{equation}
    d_{i} = \sqrt{
    \left(\frac{\theta_{i} - \hat{\theta}_{i}}{\mathrm{IQR}_{\theta}}\right)^{2} +
    \left(\frac{Zg_{i} - \hat{Zg}_{i}}{\mathrm{IQR}_{Zg}}\right)^{2} +
    \left(\frac{Zr_{i} - \hat{Zr}_{i}}{\mathrm{IQR}_{Zr}}\right)^{2}
    }
\label{eq:Euclidean_distance}
\end{equation}
in which $\mathrm{IQR}_{\theta}$, $\mathrm{IQR}_{Zg}$, and $\mathrm{IQR}_{Zr}$ are $0.113$'', $0.056$ mag, and $0.112$ mag, respectively. As displayed in Figure~\ref{fig:distribution_d}, the distribution of $d$ has a peak at $d \approx 2$ and then falls off. It also shows outliers for $d>10$, which we exclude by imposing a conservative threshold at $d=8$, to remove possible wrong cross-matches. 
This excludes 30 outliers, whose positions in the multi-dimensional space are marked in Figure~\ref{fig:verifying_crossmatch}. After the outlier removal step, we obtain 864 good cross-matches from the ZTF query results. We are aware that by using magnitudes as cross-match criteria, we may remove cataclysmic variable sources. However, this should not exclude periodic sources such as pulsating stars or binary stars, which are the main focus of this study.

We then perform Pearson's $\chi^2$ test on the obtained catalogue, similarly as in \ref{subsec:robust}, to evaluate the false positive rate (FPR) and find 284 possible false-positives ($FPR=33\%$).

\subsection{Cleaning ZTF light curves}

\label{subsec:ztf_light_cleaning}
After selecting good cross-matches, in the next step, we employ the ZTF flags recommended by \cite{GuidryEtal2021} to remove contaminated measurements in each ZTF time series. Specifically, we require good ZTF photometry measurements to follow these three criteria:
\begin{equation}
\begin{aligned}
    &\tt{catflag} = 0 \\
    &|\tt{sharp}| < 0.25 \\
    &\tt{mag} < \tt{limitmag} - 1.0. \\
\label{eq:ztf_quality_criteria}
\end{aligned}
\end{equation}
Having performed this preliminary cleaning, we then apply a $5\sigma$ clipping selection aiming to remove the potential remaining outliers. Indeed, we observed that, for some ZTF light curves, simply applying the criteria outlined in Equation~(\ref{eq:ztf_quality_criteria}) did not suffice to exclude probable outliers. \\

\subsection{Period search methodology}

\begin{figure*}
    \centering
    \begin{subfigure}{0.95\columnwidth}
    \centering
    \includegraphics[width=\linewidth]{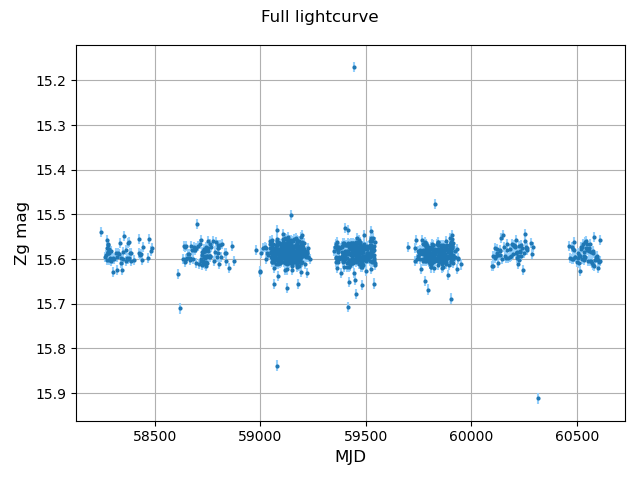}
    \caption{}
    \label{subfig:full_LC}
    \end{subfigure}
    \begin{subfigure}{0.95\columnwidth}
    \centering
    \includegraphics[width=\linewidth]{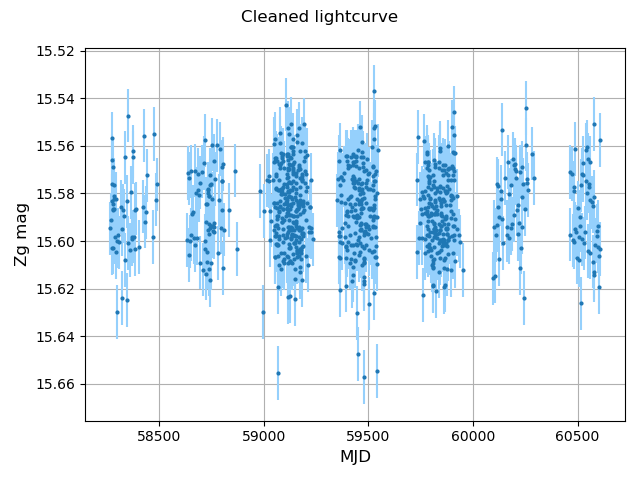}
    \caption{}
    \label{subfig:cleaned_LC}
    \end{subfigure}
    \begin{subfigure}{0.95\columnwidth}
    \centering
    \includegraphics[width=\linewidth]{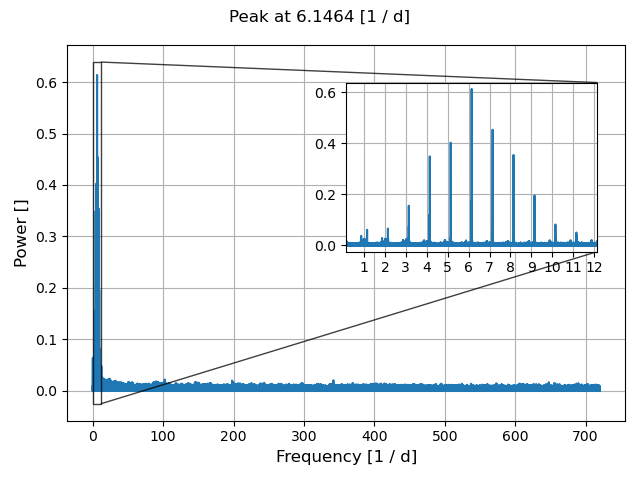}
    \caption{}
    \label{subfig:periodogram}
    \end{subfigure}
    \begin{subfigure}{0.95\columnwidth}
    \centering
    \includegraphics[width=\linewidth]{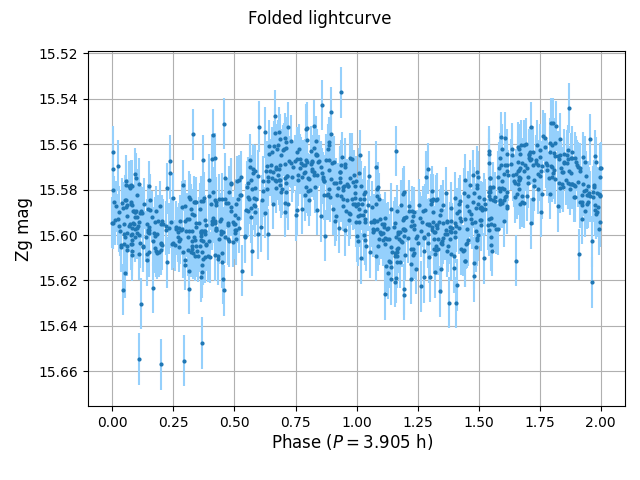}
    \caption{}
    \label{subfig:phase_folded}
    \end{subfigure}
    \caption{We illustrate the step-by-step reduction of a raw ZTF DR23 light curve (Gaia DR3 2833849800205759360) into a phase-folded light curve. 
    Between (a) and (b), outlier observations are removed from the raw data using the procedure described in Subsection~\ref{subsec:ztf_light_cleaning}. 
    In this case, 93 of the 1067 initial photometric measurements were flagged as outliers. 
    The cleaned light curve is then examined (c) using a multiband Lomb-Scargle periodogram (here with a peak at $6.1464$ cycles per day). 
    In (d) we present a folded light curve using the determined period.}
    \label{fig:reductionstepsV2}
\end{figure*}

We then analyze our cleaned time series with generalised Lomb-Scargle periodograms \citep[introduced by][]{Lomb1976,Scargle1982}. In particular, we use the \texttt{Astropy} \citep{collaboration_astropy_2018,collaboration_astropy_2022} implementation of the multi-band periodogram \citep[developed by][]{vanderplas_periodograms_2015}, scanning frequencies ranging from $f=0$ to $f=\unit[0.5]{min^{-1}}~(=\unit[720]{day^{-1}})$, with an oversampling factor $n_0=10$. 
Furthermore, we use $(N_{base},N_{band})=(1,0)$ (respectively the number of frequency terms common to both bands, and the number of frequency terms for the residuals between the base model and the individual bands), focusing our search on simple periodic variations with a common to both bands.
This choice was made as it significantly decreases computation time.
Indeed, as our targets can have long observational baselines ($\gtrsim\unit[6]{years}$), computing the periodogram with a well-defined frequency grid can require up to $N_{eval}\sim10^6$ evaluations, and adding additional base terms increases computation time by a factor of $~60$ (from $\sim\unit[1]{min}$ to $\sim\unit[1]{h}$ for $N_{base}=2$) per term.
Additionally, we compute the false-alarm probability at the extracted peak frequency in both bands (rather than in the multi-band periodogram, as it has not been implemented) using the Baluev upper limit \citep[see][]{baluev_assessing_2008}.
An illustration of this reduction process is displayed in Figure~\ref{fig:reductionstepsV2}.
We then select variable stars with a pulsating frequency $f$ by imposing the following criteria:
\begin{equation}
\begin{aligned}
    &FAP_{Zg\ \mathrm{resp.}\ Zr}(f)<10^{-3}, \\
    &FAP_{Zr\ \mathrm{resp.}\ Zg}(f)<10^{-2}, \\
    &LS(f)>0.1 \\
\end{aligned}
\label{eq:variable_select}
\end{equation}
with $FAP_{\alpha}(f)$ being the false alarm probability of a signal at frequency $f$ in the band $\alpha$, and $LS(f)$ the multi-band Lomb-Scargle power at frequency $f$. The last criterion ($LS(f)>0.1$) allows us to exclude peaks caused by very noisy signals.
Furthermore, we perform a visual check of the phase-folded light curves and periodogram in the cases where $\left|f\bmod\unit[1.00274]{day^{-1}}\right|<10^{-2}$ (i.e. if the computed frequency is close to one cycle per sidereal day, hinting towards a windowing effect from the daily observations) or $LS(f)<0.25$ (hinting towards a noisy signal). Furthermore, in the cases where $\left|f\bmod\unit[1.00274]{day^{-1}}\right|<10^{-2}$, we apply the periodogram to the \textit{Gaia} light curves (if they are available), to confirm or exclude the periodicity.\\

Out of our 864 candidates, we confirm 141 ($16\%$) periodic variables, 7 ($1\%$) undetermined variables (variable stars for which we could not determine a pulsation frequency), and 716 ($83\%$) stars for which no determined variability.
This last group's significant size, especially regarding compared to the number of expected false positives (284), can be traced back to two main reasons:
\begin{itemize}
    \item This study focuses on checking whether white dwarfs present periodic variability rather than establishing if they are variable or not, thus these stars can still present some variability without showing clear periodic signal - especially given our candidacy criteria for variability in Subsection~\ref{sub:candidates}.
    \item For stars presenting a periodic variability, our selection criteria on the Lomb-Scargle analysis, outlined in Equation~(\ref{eq:variable_select}) are possibly too harsh, in particular the Lomb-Scargle power cut $LS(f)>0.1$, as it can exclude stars with low amplitude-to-noise ratio (where the noise can induce spurious peaks and reduces the peak power value) and stars with high frequencies (which usually yield a lower peak power value due to spurious peaks).
\end{itemize}

\begin{figure*}
    \centering
    \includegraphics[width=0.8\textwidth]{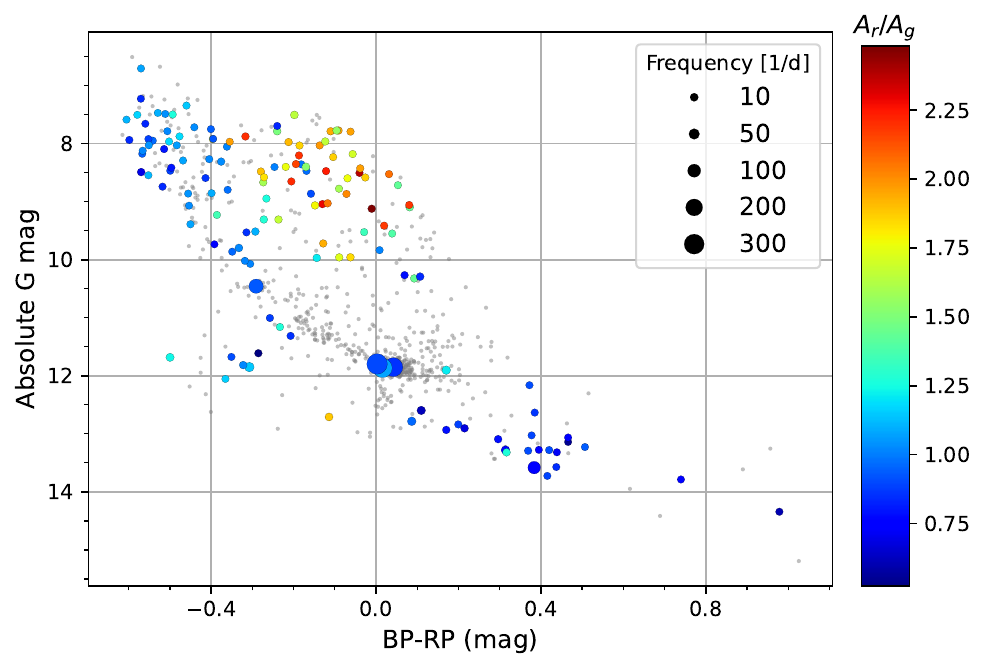}
    \caption{Position of the determined periodic stars in the HR diagram, with the colour and size of the dots representing respectively their $r/g$ amplitude ratios (see (\ref{eq:Amp_ratio})) and frequency. The background grey dots represent the non-periodic stars in our candidate list.}
    \label{fig:periodic}
\end{figure*}

Finally, the location on the HR diagram of the periodic stars is displayed in Figure~\ref{fig:periodic}, along with their frequency, and amplitude ratio $A_{r}/A_{g}$. The amplitude in the $\alpha$ band is defined as
\begin{equation}
A_{\alpha}= \mathcal{Q}_{0.95}(\alpha)-\mathcal{Q}_{0.05}(\alpha)
\label{eq:Amp_ratio}
\end{equation}
with $\mathcal{Q}_p$ the $p$-th quantile of the cleaned observations in a given band.

\section{Periodic white dwarfs}
\label{Sect:pulsating_white_dwarf_candidates}
\subsection{Classifying periodic White Dwarfs}

\begin{figure}
    \centering
    \includegraphics[width=0.95\columnwidth]{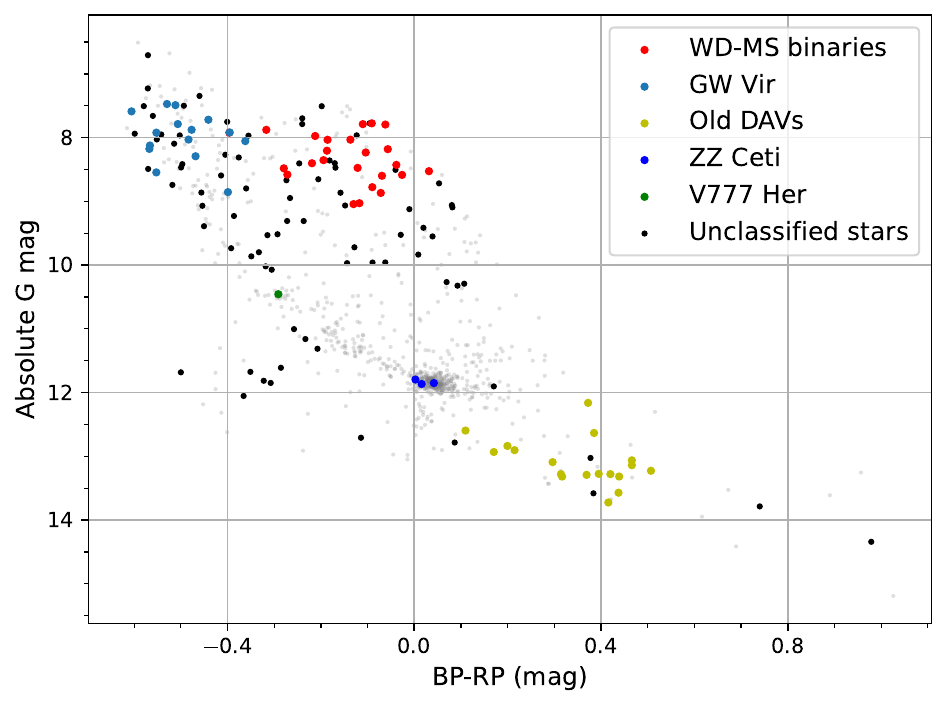}
    \caption{Position in the HR diagram of the three clusters identified using unsupervised clustering, with the manually identified ZZ Ceti and V777 Her stars added. The background grey dots are the stars we found to be non-periodic.}
    \label{fig:clustering}
\end{figure}

Following \cite{AlthausEtAL2010, CorsicoEtal2019}, the main groups of pulsating white dwarfs are the ZZ Ceti, V777 Her, GW Vir, and DQVs. Using automatic unsupervised clustering, we attempt to retrieve at least some of these groups in our sample of pulsating white dwarfs.
In particular, we use \texttt{sklearn}'s implementation of the \texttt{OPTICS} clustering algorithm \citep[see][]{ankerst_optics_1999, schubert_improving_2018} with hyperparameter $\texttt{min\_samples}=6$. The parameters used for clustering are the stars' absolute \textit{Gaia} G magnitude, \textit{Gaia} BP-RP colour, $\log_{10}$ frequency, and $\log_{10}$ amplitude ratio ($A_r/A_g$).  
We retrieve 3 clusters, totalling 82 classified and 57 unclustered stars. 
Furthermore, we perform a sanity check through visual inspection of Figure~\ref{fig:periodic} to identify possible clusters, during which we identify the same 3 clusters, a fourth smaller one, and a lone isolated star.
The first consists of stars with $A_{r}/A_{g}\gtrsim1.5$ (indicated by the yellow-red colours) and $f\lesssim\unit[50]{day^{-1}}$ lying above the white dwarf sequence (absolute G$<$10 and $\mathrm{BP - RP} \in [-0.4,0.1]$). 
These are presumed to be white dwarfs - main-sequence binary stars due to their high $Zr$ amplitude.
The second group consists of blue ($\mathrm{BP - RP} \lesssim -0.4$) stars along the main white dwarf track, with $f\lesssim\unit[50]{day^{-1}}$ and $A_{r}/A_{g}\lesssim1.25$. We identify them as possible GW Vir stars. 
The third group corresponds to old variable white dwarfs, mostly DAVs, with absolute G$>12$, $\mathrm{BP - RP} >0$, an amplitude ratio between 0.75 and 1, and $f<\unit[100]{day^{-1}}$.
The small, visually identified, cluster consists of three high-frequency ($f>\unit[200]{day^{-1}}$) stars at absolute G $\sim 11.5$ and $\mathrm{BP - RP} \sim 0.05$. These are identified as ZZ Ceti stars due to their characteristic frequencies, position in the HR diagram, and amplitude ratios between 1 and 1.25. They were not clustered by the \texttt{OPTICS} algorithm, as they are too few to be recognised as a cluster. 
This raises a discrepancy between the visible over-density of the ZZ Ceti region on Figure~\ref{fig:clustering} and the low number of identified periodic stars in the same region. We suspect this to be caused by the high frequency of ZZ Ceti stars (up to $\sim900$ cycles a day) - increasing the Lomb-Scargle periodogram's sensitivity to noise in the measurements, which decreases the peak power value of the periodogram. Furthermore, the detected ZZ Ceti stars' $Zg$ and $Zr$ amplitudes (respectively $\unit[86]{millimag}$ and $\unit[82]{millimag}$) are approximately half those of the entire dataset of periodic stars (respectively $\unit[149]{millimag}$ and $\unit[186]{millimag}$), further increasing the potential for spurious peaks in the periodograms and reducing the power value of the true frequency peak.
Nevertheless, this outlines the significant potential to identify new ZZ Cetis in this region of the HR diagram in the future.\\
Finally, similarly as for the three ZZ Ceti stars, we identify a possible V777 Herculis star in the expected clump, at absolute $\mathrm{G}=10.46$ and $\mathrm{BP - RP} = -0.22$ with a frequency $f=\unit[176.6]{day^{-1}}$, clearly in the range outlined by \cite{AlthausEtAL2010, CorsicoEtal2019} for these kind of stars.

\subsection{Stars to investigate} 
After visual inspections of the phase-folded light curves, we identified peculiar characteristics in the phase-folded light curves of at least 6 of our 141 periodic variables. 
In particular,
Gaia DR3 103999471976858496 presents a peculiar bump, indicating a secondary period, such as a transit;
Gaia DR3 1191504471436192512 and Gaia DR3 930093722208184448, both classified as WD-MS binaries, present characteristics of eclipsing binaries.
Notably, both of these are already known as eclipsing binaries, and Gaia DR3 1191504471436192512 is the already well-known NN Serpentis system \citep[see][]{NN_Ser_2010MNRAS.402.2591P}.
The remaining three, Gaia DR3 6770227729752288256, Gaia DR3 6844375121726139520, and Gaia DR3 4318508939464901760, present an intriguing "double band" feature in their Zr light curves.
The 6 phase-folded light curves are provided as an appendix (see Figure~\ref{fig:further_analysis_periodic}).
Furthermore, we recommend further investigations for the previously unstudied sources categorised as undetermined variables in our work. These are denoted by including FAR (Further Analysis Recommended) in the \textit{note} column of the data table produced with this work.

\subsection{Comparison with Literature}
\subsubsection{Crossmatch with SIMBAD}
We cross-match our catalogue of white dwarfs with the SIMBAD database \citep{2000A&AS..143....9W} to compare how many of them are already known variables.
It is important to note that, for this comparison, we use the \texttt{Object type} field provided by SIMBAD and restrain ourselves to stars classified as variables (V$^*$). 
Indeed, manually checking the literature for each of the 864 sources would prove unfeasible.
We find that out of our 864-star sample (i.e. probable variable WD identified by Gaia and have ZTF data), 49 are classified as non-cataclysmic variable stars and 18 as cataclysmic binaries (CV).
For the 148 variable stars (i.e. variable WD identified by both Gaia and ZTF photometry), we find 31 already classified as such in SIMBAD, with 13 of them classified as cataclysmic.
In detail, out of our 24 identified WD-MS binaries, 11 of them are known variables in SIMBAD (4 non-cataclysmic, 7 cataclysmic); for the GW Vir stars, one is recorded as variable out of 15; one out of the three ZZ Ceti stars; one out of 18 Old DAVs; and the lone V777 Her star is recorded as variable too. The 11 remaining known cataclysmic binary systems are unclassified in our candidate sample.

\subsubsection{Crossmatch with the MWDD}
We proceed similarly with the Montreal White Dwarf Database\footnote{\url{https://www.montrealwhitedwarfdatabase.org/home.html}} \citep[MWDD - introduced by][]{dufour_montreal_2017}, as it holds more details regarding white dwarf classes, which will help ascertain the quality of our classifications.
Out of our 864 star sample, 345 are already present in the MWDD, with 65 of the 148 variable stars already classified, and 280 of the 716 non-variable.
In particular, out of our 24 identified WD-MS binaries, 4 are already classified as such in the MWDD, for the GW Vir stars, 5 are already known out of 15, 2 of the 3 ZZ Cetis, 16 of the 18 old DAVs, and our lone V777 Her being previously classified too.\\

The result of both of these cross-matches is listed in the \textit{note} column of the data table resulting from this work (see Table~\ref{tab:excerpt} for a sample of the table). The complete table will be provided as a companion to this paper on the VizieR database\footnote{\url{https://cdsarc.u-strasbg.fr/}} \citep[introduced by][]{ochsenbein_vizier_2000}.

\section{Conclusion}
\label{Sect:Conclusion}
Using the photometric and astrometric measurements of \textit{Gaia} DR3, we propose 1423 candidates for variable white dwarfs. 
Applying a multiband Lomb-Scargle analysis on 864 stars having ZTF DR23 time series (out of 1423 initial candidates), we confirm the variability of 148 candidates, including 141 presenting periodic variability.
Using unsupervised clustering, we regroup them into 5 main groups, ZZ Ceti, GW Virginis, WD-MS binaries, Old DAVs and V777 Herculis; This yields three ZZ Ceti candidates, 15 GW Virginis candidates, 24 WD-MS binary systems, 18 Old DAVs, and one V777 Herculis.  

Furthermore, we compare our catalogue with SIMBAD and the MWDD for previously identified sources. We announce 67~newly identified periodically variable white dwarfs.
Among these, we find one potentially new ZZ Ceti star, 10 potentially new GW Virginis stars, 13 potentially new WD-MS binary systems, and two potentially new Old DAVs.  

A sample of our 864 ZTF-observed variable candidates, along with their classifications, frequencies, and amplitudes, is provided in Table~\ref{tab:excerpt}. The complete list of our candidates is provided as a companion to the paper, on the VizieR database of the Centre de Donn\'ees astronomiques de Strasbourg.

In the future, starting with \textit{Gaia} DR4 (expected by the end of 2026), photometric time series will be available in all bands (G, BP, RP, and G\_RVS) as well as in radial velocities (for the bright part of the survey). The combination of \textit{Gaia} data (astrometry, photometry, and radial velocities) with ZTF observations will enable detailed studies of white dwarf variability. In particular, aliasing issues will be mitigated thanks to the different spectral windows of the two surveys. These developments promise major progress in our understanding of variable white dwarfs.

\begin{acknowledgements}
    We thank Richard I.\ Anderson, Anthony Brown, Boris Gaensicke, Nami Mowlavi, and Giordano Viviani for discussions and comments. 
    We thank also late Prof. Gilles Fontaine for his encouragements at the beginning of this work. 
    Additionally, we thank the anonymous referee for their work and insightful comments.
    We also thank the ThinkSwiss Research Scholarship for financially support Thinh Nguyen to do research at the University of Geneva, and we thank the University of Geneva for their welcome and hospitality to the guest student.
    This work has made use of data from the European Space Agency (ESA) mission \textit{Gaia} (\url{https://www.cosmos.esa.int/gaia}), processed by the \textit{Gaia} Data Processing and Analysis Consortium (DPAC, \url{https://www.cosmos.esa.int/web/gaia/dpac/consortium}). 
    Funding for the DPAC has been provided by national institutions, in particular the institutions participating in the \textit{Gaia} Multilateral Agreement.
    This work is partly based on observations obtained with the Samuel Oschin Telescope 48-inch and the 60-inch Telescope at the Palomar Observatory as part of the Zwicky Transient Facility project. 
    ZTF is supported by the National Science Foundation under Grants No. AST-1440341 and AST-2034437 and a collaboration including current partners Caltech, IPAC, the Oskar Klein Center at Stockholm University, the University of Maryland, University of California, Berkeley , the University of Wisconsin at Milwaukee, University of Warwick, Ruhr University, Cornell University, Northwestern University and Drexel University. 
    Operations are conducted by COO, IPAC, and UW.
    This reaserch also made use of; the NASA/IPAC Infrared Science Archive, which is funded by the National Aeronautics and Space Administration and operated by the California Institute of Technology; the VizieR catalogue access tool and the SIMBAD Astronomical Database, both operated by the CDS, Strasbourg Astronomical Observatory, France; and the Montreal White Dwarf Database.
    
\end{acknowledgements}

\bibliographystyle{aa}
\bibliography{biblio}

\appendix
\section{Additional materials}

\counterwithin{figure}{section}

\begin{figure*}[h]
    \centering
    
    \begin{subfigure}{0.85\columnwidth}
    \centering
    \includegraphics[width=\linewidth]{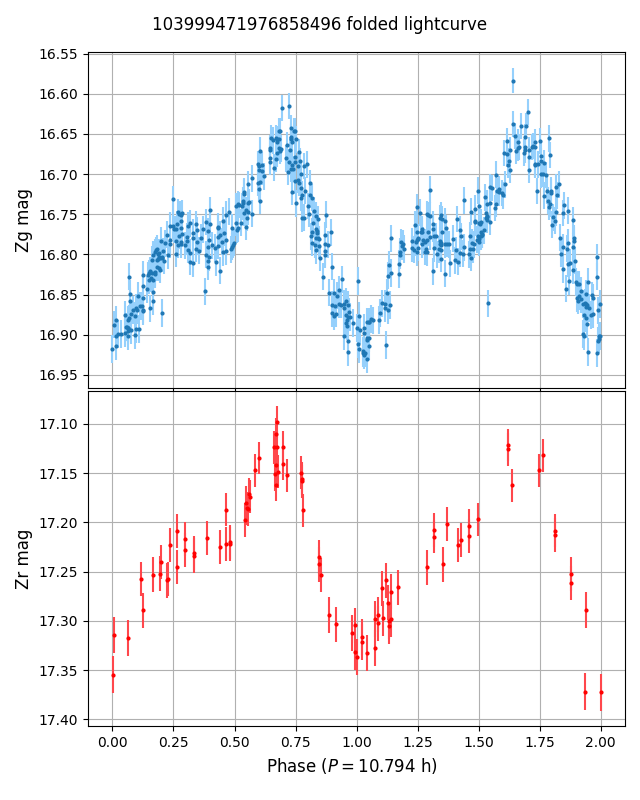}
    \caption{Gaia DR3 103999471976858496}
    \label{fig:103999471976858496}
    \end{subfigure}
    \hfill
    \begin{subfigure}{0.85\columnwidth}
    \centering
    \includegraphics[width=\linewidth]{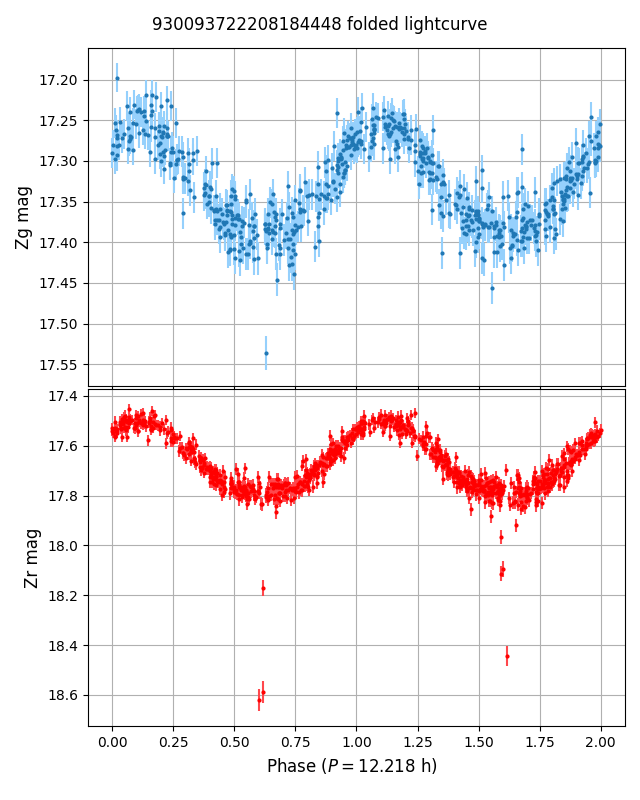}
    \caption{Gaia DR3 930093722208184448}
    \label{fig:930093722208184448}
    \end{subfigure}
    \hfill
    \begin{subfigure}{0.85\columnwidth}
    \centering
    \includegraphics[width=\linewidth]{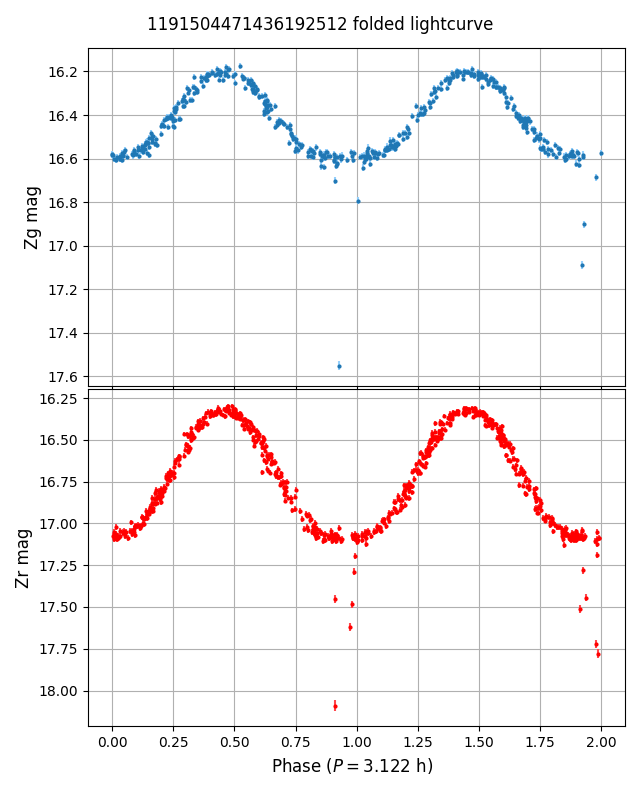}
    \caption{Gaia DR3 1191504471436192512}
    \label{fig:1191504471436192512}
    \end{subfigure}
    \hfill
    \begin{subfigure}{0.85\columnwidth}
    \centering
    \includegraphics[width=\linewidth]{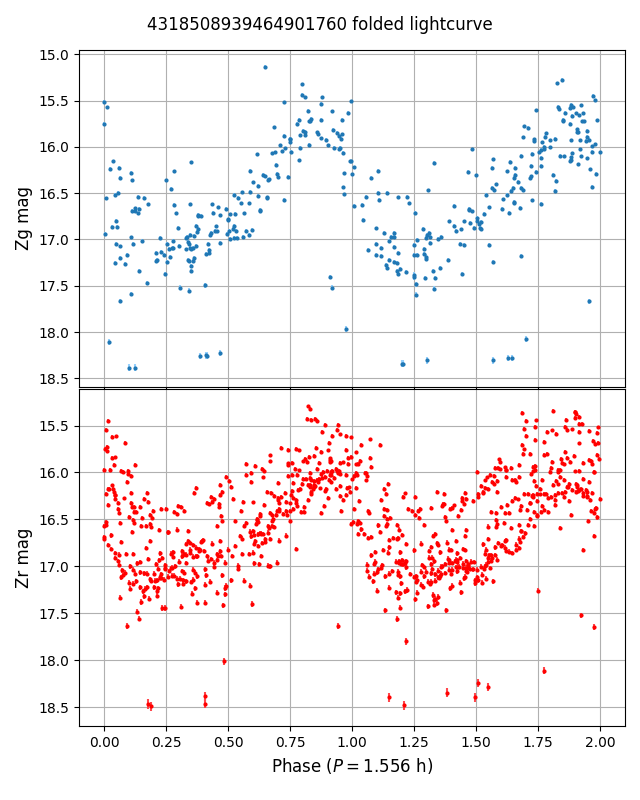}
    \caption{Gaia DR3 4318508939464901760}
    \label{fig:4318508939464901760}
    \end{subfigure}
    \hfill
    
\end{figure*}

\begin{figure*}[h]\ContinuedFloat
    \centering
    
    \begin{subfigure}{0.85\columnwidth}
    \centering
    \includegraphics[width=\linewidth]{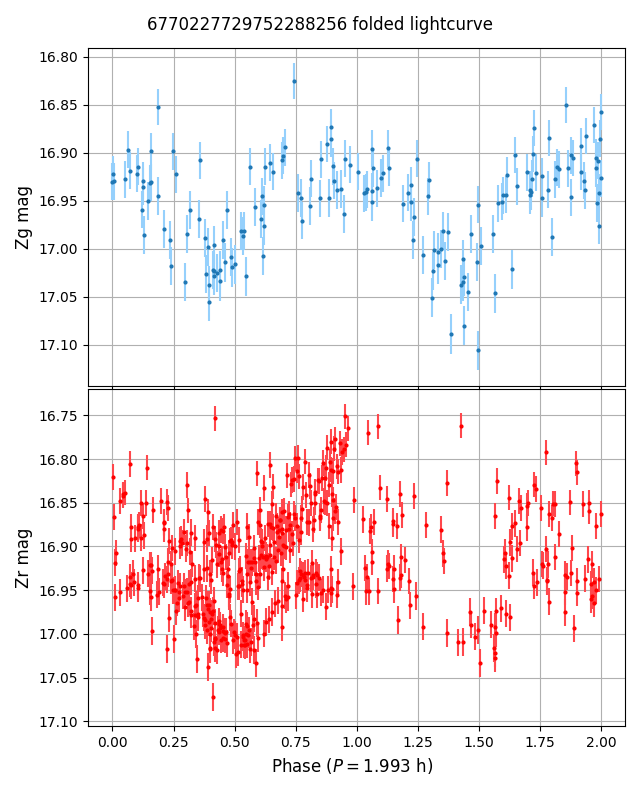}
    \caption{Gaia DR3 6770227729752288256}
    \label{fig:6770227729752288256}
    \end{subfigure}
    \hfill
    \begin{subfigure}{0.85\columnwidth}
    \centering
    \includegraphics[width=\linewidth]{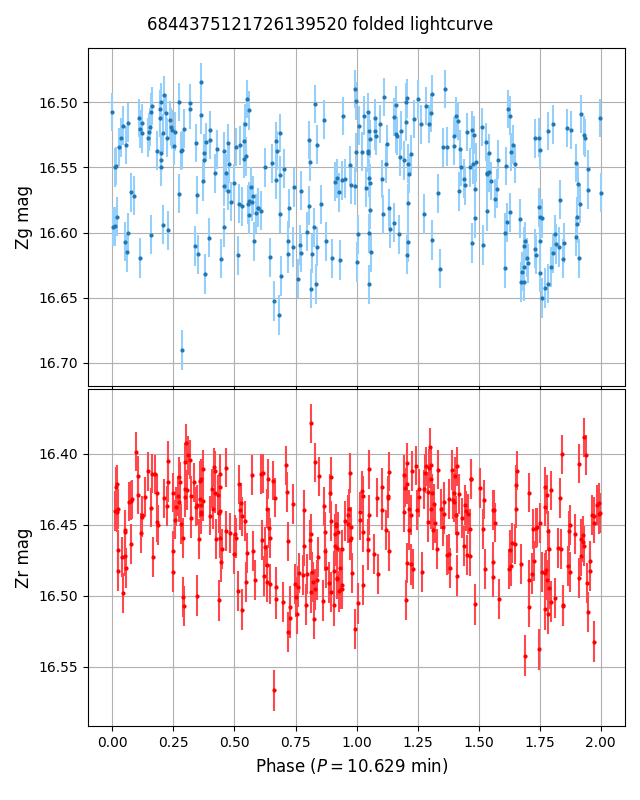}
    \caption{Gaia DR3 6844375121726139520}
    \label{fig:6844375121726139520}
    \end{subfigure}
    \hfill
        
    \caption{Phase-folded light curves of the 6 periodic sources warranting further investigations.}
    \label{fig:further_analysis_periodic}
\end{figure*}
\begin{table*}
\centering
\caption{Excerpt of the companion table, the frequencies are reported in [day$^{-1}$]}
\begin{tabular}{cccccccccc}
\hline \hline
WDJname & GaiaDR3\_id & ra & e\_ra & dec & e\_dec & plx & e\_plx & G\_mag & RP\_mag \\
\hline
WDJ135309.97+484021.17 & 1510467090935595008 & 208.29 & 0.04 & 48.67 & 0.04 & 5.7 & 0.05 & 16.68 & 16.88 \\
WDJ052038.32+304823.92 & 3446909137068558464 & 80.16 & 0.04 & 30.81 & 0.02 & 17.51 & 0.04 & 15.58 & 15.59 \\
WDJ111026.19+191229.75 & 3984115430179696128 & 167.61 & 0.09 & 19.21 & 0.07 & 16.65 & 0.1 & 17.17 & 16.91 \\
WDJ070223.05-174931.25 & 2935237657195591424 & 105.6 & 0.04 & -17.83 & 0.05 & 1.81 & 0.06 & 16.51 & 16.55 \\
WDJ220247.69+275010.67 & 1893101535448502400 & 330.7 & 0.05 & 27.84 & 0.07 & 2.23 & 0.07 & 16.55 & 16.86 \\
WDJ105010.80-140436.76 & 3750072904055666176 & 162.54 & 0.08 & -14.08 & 0.06 & 9.15 & 0.09 & 17.14 & 17.09 \\
WDJ143406.77+150817.81 & 1228266814506156928 & 218.53 & 0.04 & 15.14 & 0.03 & 12.68 & 0.04 & 16.01 & 16.09 \\
WDJ024849.59+264332.15 & 114808397128552576 & 42.21 & 0.11 & 26.73 & 0.15 & 5.67 & 0.11 & 17.23 & 17.34 \\
WDJ163914.29+474835.84 & 1410345596469085184 & 249.81 & 0.07 & 47.81 & 0.07 & 6.71 & 0.08 & 17.76 & 17.78 \\
WDJ022917.15-150350.42 & 5146019876066016000 & 37.32 & 0.08 & -15.06 & 0.08 & 5.48 & 0.1 & 17.71 & 17.63 \\
WDJ220420.60+282321.36 & 1893512958955407232 & 331.09 & 0.08 & 28.39 & 0.08 & 5.27 & 0.1 & 17.74 & 17.67 \\
\hline
\end{tabular}
\label{tab:excerpt}
\end{table*}

\begin{table*}\ContinuedFloat
\begin{tabular}{cccccccccl}
\hline \hline
BP\_mag & g\_mag & r\_mag & Variable & Periodic & frequency & g\_amp & r\_amp & classification & note \\
\hline
16.59 & 16.54 & 16.92 & True & True & 176.58 & 0.11 & 0.1 & V777 Her & in SIMBAD (V*), in MWDD \\
15.59 & 15.53 & 15.73 & True & True & 320.01 & 0.07 & 0.06 & ZZ Ceti & in SIMBAD (V*), in MWDD \\
17.33 & 17.35 & 17.2 & True & True & 5.38 & 0.13 & 0.12 & Old DAVs &  \\
16.44 & 16.49 & 16.89 & True & True & 8.82 & 0.28 & 0.55 & WD-MS binaries &  \\
16.39 & 16.38 & 16.89 & True & True & 1.23 & 0.06 & 0.06 & GW Vir & in SIMBAD (V*) \\
17.2 & 17.09 & 17.17 & True & False & 0.0 & 0.31 & 0.48 &  & in SIMBAD (CV) \\
15.99 & 15.9 & 16.17 & False & False & 0.0 & 0.0 & 0.0 &  & in MWDD \\
17.19 & 17.09 & 17.39 & False & False & 0.0 & 0.0 & 0.0 &  &  \\
17.81 & 17.72 & 17.9 & False & False & 0.0 & 0.0 & 0.0 &  &  \\
17.8 & 17.71 & 17.79 & False & False & 0.0 & 0.0 & 0.0 &  &  \\
17.86 & 17.77 & 17.86 & False & False & 0.0 & 0.0 & 0.0 &  &  \\
\hline
\end{tabular}
\end{table*}


\end{document}